\documentclass[preprint, 5p, twocolumn, 10pt]{elsarticle}
 
\usepackage{amssymb}
\usepackage{xspace}
\usepackage{color}
\usepackage[dvipsnames]{xcolor}
\definecolor{MatlabCellColour}{RGB}{235,235,235}
\definecolor{LammpsCodeColor}{RGB}{223,238,247}
\newcommand{\pl}{(py)LIon\xspace}
\newcommand{\ca}{$^{40}\text{Ca}^{+}$\xspace}
\newcommand{\speciesNH}{$\text{NH}_3^+$}
\newcommand{\eg}{\textit{e.g.~}}

\newcommand{\code}[1]{\colorbox{MatlabCellColour}{\texttt{#1}}}
\newcommand{\lammps}[1]{\colorbox{LammpsCodeColor}{\texttt{#1}}}
\newcommand{\lammpsSoftware}{LAMMPS\xspace}
\newcommand{\rf}{rf}
\newcommand{\timestep}{timestep}

\usepackage{xstring}
\usepackage{suffix}
\newcommand*{\refform}[1]{%
	\IfBeginWith{#1}{eq:}{Eq.~\eqref{#1}}{}%
	\IfBeginWith{#1}{fig:}{Figure~\ref{#1}}{}%
	\IfBeginWith{#1}{tab:}{Table~\ref{#1}}{}%
	\IfBeginWith{#1}{appendix:}{Appendix~\ref{#1}}{}%
	\IfBeginWith{#1}{sec:}{Section~\ref{#1}}{}%
}
\newcommand*{\coordAxis}[1]{\ensuremath{\mathbf{\hat{e}}_#1}}
\usepackage{tabularx}

\usepackage{amsmath}

\usepackage{url}
\usepackage{epstopdf}
\usepackage{siunitx}

\usepackage{listings}
\usepackage{setspace} 
\usepackage{booktabs}
\definecolor{mygreen}{RGB}{28,172,0} 
\definecolor{mylilas}{RGB}{170,55,241}

\newcounter{bla}

\journal{Computer Physics Communications}

\begin{document}

\begin{frontmatter}



\title{\pl: a package for simulating trapped ion trajectories}


\author[ox]{E. Bentine\corref{author}}
\author[ox]{C. J. Foot}
\author[ino]{D. Trypogeorgos}

\cortext[author] {Corresponding author.\\\textit{E-mail address:} elliot.bentine@physics.ox.ac.uk}
\address[ox]{Clarendon Laboratory, Department of Physics, University of Oxford, Parks Road, Oxford, OX1 3PU, UK}
\address[ino]{INO-CNR BEC Center and Dipartimento di Fisica, Universit\'a di Trento, 38123 Povo, Italy}

\begin{abstract}
The \pl package is a set of tools to simulate the classical trajectories of ensembles of ions in electrodynamic traps.
Molecular dynamics simulations are performed using \lammpsSoftware{}, an efficient and feature-rich program.
\pl has been validated by comparison with the analytic theory describing ion trap dynamics.
Notable features include GPU-accelerated force calculations, and treating collections of ions as rigid bodies to enable investigations of the rotational dynamics of large, mesoscopic charged particles.
\end{abstract}


\end{frontmatter}



{\bf PROGRAM SUMMARY}

\begin{small}
\noindent
{\em Manuscript Title:} \pl: a package for simulating trapped ion trajectories \\
{\em Authors:} E. Bentine, C. J. Foot, D. Trypogeorgos                                               \\
{\em Program Title:}  \pl                                        \\
{\em Licensing provisions:} MIT                                  \\
{\em Programming language:} Matlab, Python                                 \\
{\em Computer:} pc, cluster                                              \\
{\em Operating system:} Windows, Linux, Mac                                    \\
{\em RAM:} size-dependent                                            \\
{\em Number of processors used:} user-configurable                              
{\em Classification:} 2 Atomic Physics, 12 Gases and Fluids, 16 Molecular Physics and Physical Chemistry                                       \\
{\em Subprograms used:} LAMMPS                                      \\
{\em Nature of problem:} 
Simulating the dynamics of ions and mesoscopic charged particles confined in an electrodynamic trap using molecular dynamics methods
   \\
{\em Solution method:} 
Provide a tested, feature-rich API to configure molecular dynamics calculations in LAMMPS
   \\
{\em Unusual features:} 
\pl can treat collections of ions as rigid bodies to simulate larger objects confined in electrodynamic traps. GPU acceleration is provided through the LAMMPS \lammps{gpu} package.
   \\
{\em Running time:}
Size-dependent, ranges from seconds to hours on a recent workstation.
\end{small}

\section{Introduction}
\label{sec:Introduction}

Electrodynamic ion traps, also known as Paul traps, are widely used in physics and chemistry to confine charged particles~\cite{Paul1990}.
Their applications include mass spectrometry, quantum chemistry, ultra-high precision frequency measurements, and quantum computation~\cite{marchQuadrupoleIonTrap2005,staffordRecentImprovementsAnalytical1984,IonTraps1996,benhelmFaulttolerantQuantumComputing2008}.
The dynamics of ions confined in these traps can be simulated using molecular dynamics techniques~\cite{vangorpSimbucaUsingGraphics2011,dahlSIMIONPCPS21990,schillerMolecularDynamicsSimulation2003}, in which classical equations of motion for the ions are integrated.

Molecular dynamics techniques are also widely used in chemistry to simulate the interactions and properties of macromolecules, membranes, polymers and other systems by computing the trajectories of the constituent atoms.
A number of programs have been developed to simulate these systems~\cite{trottGeneralPurposeMolecular2010,mattheyPROTOMOLObjectOrientedFramework2002,NAMD,Gromacs}, which typically comprise thousands to millions of atoms and are dominated by short-range forces.
These codes are efficient and feature-rich, with numerous accelerated methods for integration and both long- and short-ranged force calculations.

The \pl package is a set of tools to simulate the classical trajectories of ions in electrodynamic traps.
The time-integration is performed using \lammpsSoftware{}\footnote{Large-scale Atomic/Molecular Massively Parallel Simulator}, an established classical molecular dynamics code, which is more typically used for biological and materials modelling~\cite{trottGeneralPurposeMolecular2010}.
\pl{} offers a robust way to author ion trap simulations, with a simplified workflow that is geared towards the atomic physics community.
In addition, the results from \pl{} have been verified by comparison to analytical descriptions of ion trap dynamics.

Performing simulations in \lammpsSoftware{} allows \pl{} to leverage a number of advanced methods, models and features.
For instance, calculations can be performed using a graphics processing unit (GPU), which decreases the time taken to calculate the Coulomb repulsion between large numbers of ions~\cite{vangorpSimbucaUsingGraphics2011}.
Additionally, \pl supports fixing individual particles together to create rigid bodies.
This feature enables the rotational dynamics of trapped ions to be simulated, which is particularly interesting for systems of larger ions~\cite{Hesse2002,kaneLevitatedSpinningGraphene2010a,Delord2017}.
Such investigations are not possible in other ion trap simulation packages, which are restricted to point-like particles. 

This paper describes the use and implementation of \pl{} as follows.
In \refform{sec:dynamics}, we describe the dynamics of charged particles in electrodynamic traps.
In \refform{sec:overviewpl}, we provide an overview of \pl{}, and describe the features of a typical simulation.
In \refform{sec:example}, we walk through example scripts and benchmark the performance of \pl{}.
In \refform{sec:Validation}, we validate the output of \pl{} by comparing test simulations to analytic results.
In \refform{sec:Implementation}, we provide details of \pl{}'s implementation and discuss the configuration of \lammpsSoftware{}.
We conclude in \refform{sec:Conclusion}, and provide information for getting started in \refform{sec:gettingStarted}.

\pl{} is available for both Matlab and Python.
The workflow is similar for both languages, and for brevity we describe only the Matlab package in detail here.

\section{Dynamics of trapped ions}
\label{sec:dynamics}

Electrodynamic traps confine ions using a combination of dc and ac electric fields that exert an oscillating, position-dependent force on a charged particle to circumvent Earnshaw's theorem and allow trapping~\cite{footAtomicPhysics2004,berkelandMinimizationIonMicromotion1998}. 
An ion of charge $Q$ and mass $M$ in a linear Paul trap is confined by the quadrupole electric field
\begin{equation}
\mathbf{E}(\mathbf{x}, t) = \frac{V_0}{R_0^2}\cos\Omega t \left(x \coordAxis{x} - y \coordAxis{y}\right) + \frac{\kappa U_0}{Z_0^2}(2z\coordAxis{z} - x \coordAxis{x} - y \coordAxis{y}),
\label{eq:field}
\end{equation}
where $V_0$, $U_0$ are the amplitudes of the ac and dc voltages, $R_0$, $Z_0$ are characteristic lengths along the radial and axial directions that are related to the distance between the trap electrodes, and $\kappa$ is a geometrical factor~\cite{willitschChemicalApplicationsLaser2008,berkelandMinimizationIonMicromotion1998}.
The quantities \coordAxis{i} indicate unit vectors along the Cartesian axes.
The ion's motion in each dimension $m$ separates into a rapid `micromotion' at frequency $\Omega$ and a slow oscillation at the secular frequency $\omega_m$.
The pseudopotential approximation is often used, in which one neglects the micromotion and treats the ion as being trapped in an effective harmonic potential with secular frequency $\omega_m/2\pi$, where
\begin{equation}
\label{eq:pseudopot}
\omega_m\simeq\frac{\Omega}{2}\sqrt{a_m + \frac{1}{2}q_m^2}.
\end{equation}
The $a_m$ and $q_m$ are the Mathieu coefficients, defined as
\begin{equation}
\begin{split}
a_x&=a_y=-\frac{1}{2}a_z=-\frac{4Q\kappa U_0}{MZ_0^2\Omega^2}, \\
q_x&=-q_y=\frac{2QV_0}{MR_0^2\Omega^2}, \quad q_z=0.
\label{eq:Mathieuaq}
\end{split}
\end{equation}
The pseudopotential approximation is strictly valid only for $\vert q_m \vert,\,\vert a_m \vert \ll 1$, but it remains accurate to 1\% up to $q=0.4$.

An ensemble of trapped ions experience both the external trapping fields and a mutual Coulomb interaction. 
The total potential energy in the pseudopotential approximation is
\begin{eqnarray}
V(\vec x) = \frac{1}{2}\sum_{i=1}^NU_i(\vec x) +
\sum_{\substack{i,j=1\\i\neq j}}^N\frac{1}{8\pi\epsilon_0}\frac{Q_i Q_j}{|\vec x_j-\vec x_i|},
\label{eq:fullPotential}
\end{eqnarray}
where $U_i(\vec x) = M_i (\vec \omega_i^2 \cdot \vec x_i^2) / 2 $ is the confining potential for ions $i,j$ of mass $M_i$, secular frequencies $\vec\omega_i/2\pi= \lbrace \omega_x, \omega_y, \omega_z \rbrace / 2\pi$ and charge $Q_i$. $\epsilon_0$ is the permittivity of free space, and $N$ the number of ions. 
Equation~\ref{eq:fullPotential} excludes effects such as heating arising from the micromotion of the ions, which require a treatment using the full time-dependent electric field of \refform{eq:field}.

The ground state configuration of the system is an ordered structure called a Coulomb crystal. 
The system crystallises at sufficiently low temperatures where the kinetic energy of the ions is negligible compared to the trapping potential and the repulsive Coulomb forces, which fix the positions of the ions.
This is achieved in practice using methods such as laser cooling or buffer gas cooling to extract kinetic energy from the ion ensemble.

\section{Overview of \pl}
\label{sec:overviewpl}

\begin{figure}[ht]
  \centering
     \includegraphics[width=\columnwidth]{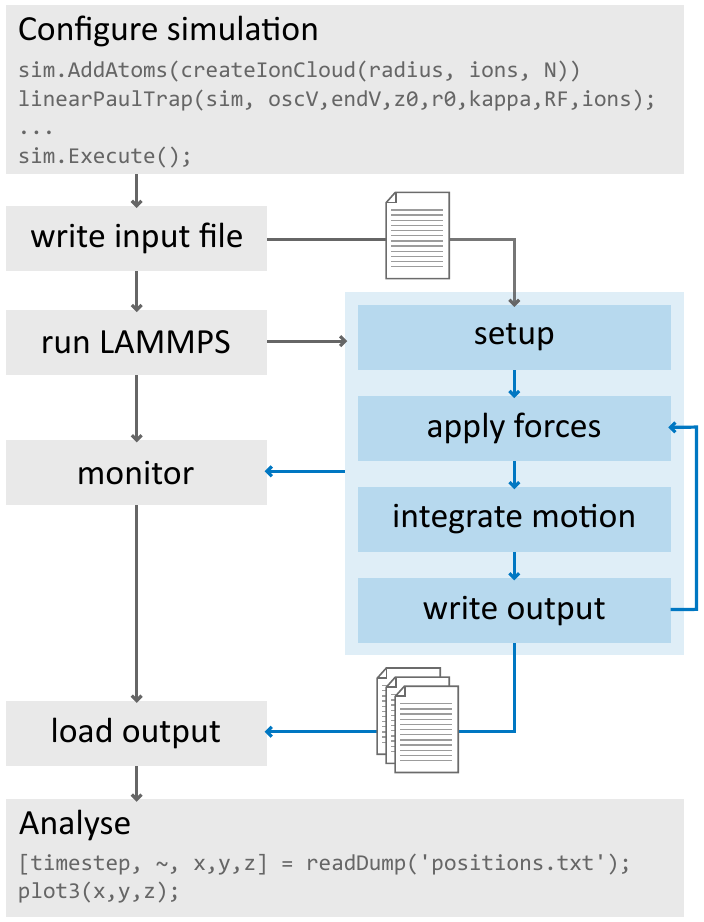}
\caption{Flow diagram depicting configuration and execution of a LAMMPS simulation using \pl. 
A simulation is defined through high-level commands in the chosen environment (grey). 
When the simulation is executed an input file to configure LAMMPS is generated and a LAMMPS child process spawned (blue).
An asynchronous loop monitors progress and checks for errors.
Time-dependent properties of the simulation are written to output file(s). 
On completion control reverts back to the main environment for post-processing and analysis.}
\label{fig:flowDiagram}
\end{figure}

\pl provides high-level functions to configure and execute simulations of charged particles in ion traps.
\refform{fig:flowDiagram} shows the process flow of a typical program.
\pl translates a configured simulation into an \emph{input file} and submits it to \lammpsSoftware{}, which performs the molecular dynamics.
Once the simulation is complete, process control reverts to the chosen environment for post-processing and analysis.

The configuration of a typical simulation in \pl proceeds as follows:
\begin{enumerate}
\item Create the simulation and specify the domain.
\item Define the atomic species used and create ions.
\item Configure the trap parameters, using either the full electric fields or the pseudopotential approximation.
\item Configure other applied forces, e.g. laser cooling, bias electric fields, Langevin baths.
\item Select which parameters to save, and the periodicity with which to save them.
\item Run the simulation to integrate the equations of motion.
\end{enumerate}

For clarity, commands and syntax in \pl{} are shown \code{like so}.

\subsection{Creating the simulation}

An instance of the \code{LAMMPSSimulation} class represents a simulation, which we refer to by the shorthand \code{sim}.
High-level \pl commands modify this object, which contains all the information required to generate the \lammpsSoftware{} input file. 
The \code{sim} object has a few global configuration properties.
The \code{NeighbourList} style and the \code{CoulombCutoff} and \code{NeighbourSkin} distances (see \refform{sec:pairwise}) configure the pairwise interactions.
The \code{GPUAccel} switch enables or disables acceleration using a GPU (see \refform{sec:GPU}).
\code{TimeStep} sets the step duration used for time evolution of the system.
When unset, \pl attempts to select an appropriate step size by considering the fastest timescale of the simulation, which is typically the period of the oscillating quadrupole field.

\subsection{Simulation domain}

\code{sim.SetSimulationDomain} defines a volume of space within which the ion trajectories are integrated.
This must be specified before adding ions or forces to the simulation. 
\pl will expand the simulation domain during time evolution when required, but the specified domain must contain all ions at the start of the simulation.

\subsection{Adding ions}

\code{sim.AddAtomType(charge, mass)} adds a new atomic species, defined by its charge in units of elementary charge $+e$, and mass in atomic mass units.
Once a species is defined, ions can be placed in the simulation. 
For instance, the command \code{atomCloud(sim, radius, species, N)}
creates \code{N} ions of the desired \code{species} randomly placed within the stated \code{radius}.
For more precise control over the ion positions, such as when defining ions to form a rigid body, \code{placeAtoms(sim, atomType, x, y, z)} inserts ions at the coordinates specified by the column vectors \code{x}, \code{y}, \code{z}.
Each ion is assigned a unique ID when placed.

Groups of ions can be specified using either the constituent atomic species or the unique IDs of the ions.
This allows fields and forces to be applied to specific collections of ions.
Setting the \code{Rigid} parameter of the group to true fixes the relative positions of ions in the group, so binding them together as a rigid body.

\subsection{Forces and fields}

\subsubsection{Electric fields and confinement}

\code{efield(Ex, Ey, Ez)} creates a static, uniform electric field of the form $\vec{E} = E_x \coordAxis{x} + E_y \coordAxis{y} + E_z \coordAxis{z}$.
Linear Paul traps are defined using \code{linearPT} (for full fields) or \code{linearPseudoPT} (for the pseudopotential approximation), with a parameterization of the fields as in~
\refform{eq:field}.
Multi-species simulations that use the pseudopotential approximation must define a separate \code{linearPseudoPT} for each charged atomic species because the pseudopotential trap frequencies depend on the specific charge $Q / M$.

\subsubsection{Cooling the motion of ions}

\pl offers two solutions to simulate cooling: coupling to a Langevin bath, and laser cooling.

\code{langevinBath(T, dampingTime, group)} creates a Langevin bath with temperature \code{T} in Kelvin and \code{dampingTime} equal to the $1/e$ velocity-relaxation time in seconds.
The optional argument \code{group} allows the bath to be applied to a chosen group, which permits \eg{} species-selective cooling or heating of ions.
The bath applies both a damping force to reduce the kinetic energy of the ions and stochastic kicks so that ions thermalise at the temperature of the bath~\cite{zhangMoleculardynamicsSimulationsCold2007a}.

\code{laserCool(species, gamma)} provides an anisotropic cooling mechanism for a particular ion \code{species}, by only damping motion parallel to the given direction $\vec{\gamma}$.
We model laser cooling as a viscous force $f_\text{laser} = - M (\vec{\gamma} \cdot \vec{v}) \vec{\gamma}$ along the direction of an applied beam that is proportional to the velocity $\vec{v}$ of each ion~\cite{zhangMoleculardynamicsSimulationsCold2007a}.
Our implementation does not exert a stochastic force and so the temperature is reduced to zero.

\subsection{Minimisation}
\label{sec:Minimisation}

The energy of a cloud of randomly placed ions comes from a combination of the inter-ion Coulomb repulsion, the potential energy in the trap, and the kinetic energy of the ions.
The \code{minimize} command uses a modified time-integration method that is well-suited for driving the system to its minimum energy configuration.
This method damps the equations of motion to extract energy and also limits the distance that each ion can move in a single time step.

Strong damping generally affects the strength of confinement experienced by ions in a Paul trap~\cite{nasseInfluenceBackgroundPressure2001,hasegawaDynamicsSingleParticle1995}.
Using \code{minimisation} with the full oscillating electric fields of the \code{linearPT} command will therefore lead to an incorrect minimum energy configuration, because the damped minimisation algorithm changes the effective harmonic trap frequencies.
To avoid this effect, minimisation should only be performed when using the \code{linearPseudoPT} confinement, which uses fixed harmonic trap frequencies that are unaffected by the presence of damping.

\subsection{Time evolution}

\code{evolve(N)} advances the simulation by integrating the Newtonian equations of motion for \code{N} steps.
By interleaving \code{evolve} with other commands it is possible to add or remove forces and fields at specific times in the simulation, creating complex sequences where trap parameters are varied in the time-domain.
An example of this is given in \refform{sec:sympatheticCooling}.

\subsection{Writing/reading data}

The \code{dump(filename, variables, steps)} command writes time-dependent per-ion variables to the given \code{filename} at regular intervals every \code{steps} timesteps. 
The cell array \code{variables} contains a mixture of string literals and/or \code{LAMMPSVariable}s, which are the internal representation of a variable in \pl.
The command \code{timeAvg(variables, duration)} creates a \code{LAMMPSVariable} that represents a time-average of another variable. For example, \code{timeAvg('vx', 1/RF)} averages the velocity in the \coordAxis{x}-direction over an \rf{} period.

Literals are a shorthand representation for selecting the properties of ions to be output from the simulation and can be any of the following types: 
\begin{enumerate}
\item ion coordinates: \code{x}, \code{y}, \code{z}
\item ion velocities: \code{vx}, \code{vy}, \code{vz}
\item the \code{id} of each ion
\item time-independent ion properties: \code{mass}, \code{charge}
\end{enumerate}

The order of ions in the output file may change as they move between processor partitions during the simulation.
To enable the reordering of data during postprocessing, the \code{dump} command automatically configures output files to include ion \code{id}s.
These are used by the \code{readDump} command to reconstruct trajectories when loading the output file.

\subsection{Execution}

\code{sim.Execute()} runs a configured simulation.
\pl generates the LAMMPS input file, launches the LAMMPS executable, and monitors its progress via an asynchronous update loop that handles errors and provides diagnostic information\footnote{A useful list of LAMMPS error codes can be found in the documentation\ \cite{LammpsManual}}.
The process terminates automatically after completion of a simulation, or if the user aborts a \pl{} simulation.
\pl{} will raise an exception if the \code{sim} object is modified after execution.
This guarantees that the state of \code{sim} remains a faithful description of the executed simulation.

\section{Examples}
\label{sec:example}

\subsection{Coulomb crystal of a single species}

We define the electric fields of an ion trap, confine a small cloud of calcium ions, and cool them into a Coulomb crystal by coupling them to a Langevin bath.

\begin{lstlisting}
sim = LAMMPSSimulation();
sim.SetSimulationDomain(1e-3,1e-3,1e-3);

caIons = sim.AddAtomType(1, 40);
createIonCloud(sim, 1e-3, caIons, 100, 1);

RF = 3.85e6;
sim.Add(linearPT(300, 7, 2.75e-3, 3.5e-3, 0.3, RF));

T = 1e-3;
sim.Add(langevinBath(T, 10e-6, sim.Group(caIons)));

sim.Add(dump('pos.txt', {'id', 'x', 'y', 'z'}, 1));
secularVel = timeAvg({'vx', 'vy', 'vz'}, 1/RF);
sim.Add(dump('secV.txt', {'id', secularVel}));

sim.Add(evolve(20000));
sim.Execute();
\end{lstlisting}

First, we create the \code{LAMMPSSimulation} instance that will represent the simulation and configure a
$\SI{1}{\milli\metre}\times\SI{1}{\milli\metre}\times\SI{1}{\milli\metre}$
initial simulation domain for the ions.
We define the \ca species with $M=\SI{40}{amu}$ and charge $Q=\SI{+1}{e}$.
A cloud of 100 randomly-placed ions is added to the simulation domain with \code{createIonCloud}.
We add a linear Paul trap with an oscillating voltage of \SI{300}{\volt}, endcap voltage of \SI{7}{\volt}, $r_0 = \SI{2.75}{\milli\m}$, $z_0 = \SI{3.5}{\milli\m}$, geometric constant $\kappa = 0.3$ and frequency $\Omega/2\pi = \SI{3.85}{\mega\Hz}$.

To cool the translational motion of the ions, we couple them to a Langevin bath that damps the velocity of the ions with a relaxation time of \SI{10}{\micro\s}, and also delivers a small random kick to each ion every time-step such that the temperature of the ensemble relaxes to \SI{1}{\milli\K}.  

\begin{figure}[ht]
\centering
     \includegraphics[width=\columnwidth]{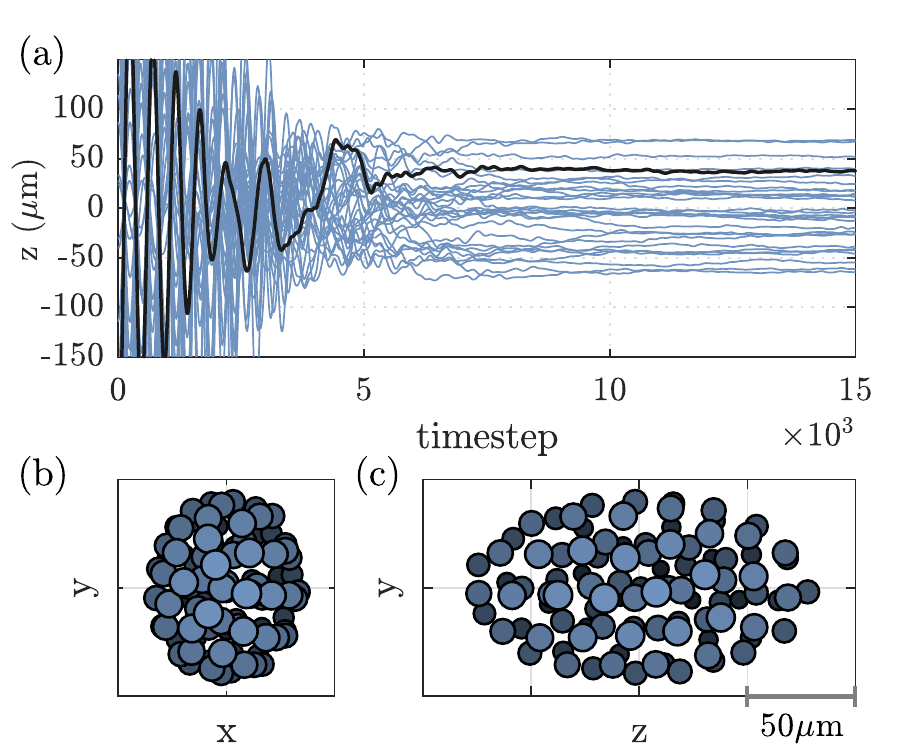}
\caption{The formation of a 100-ion Coulomb crystal. Ions start from random positions at a high temperature. A structure forms as their motion is cooled by the action of the Langevin bath over many time-steps. (a) Time evolution of the positions of ions along \coordAxis{z}, showing a gaseous phase of random, changing positions that transitions to an ordered structure. For clarity, only 30 ions are shown, and one trajectory is highlighted in black. (b) \& (c) Front and side views of the final positions of ions, illustrating the ordered Coulomb crystal structure. The shading indicates distance from the camera, with ions that are further away rendered in darker colours.}
\label{fig:exampleSim}
\end{figure}

We configure \code{sim} to output data using \code{dump}. 
In this example we output the positions of ions every \timestep{} and time-averaged secular velocities every period of the oscillating field, $2\pi/\Omega$.
The command \code{evolve} instructs the simulation to integrate for 20000 steps.
Finally, the configured \code{sim} is executed, which writes the input file, launches the LAMMPS process and runs the simulation.
The resulting trajectories and equilibrium positions are shown in Figure~\ref{fig:exampleSim}.

\subsection{Sympathetic cooling}
\label{sec:sympatheticCooling}

\pl supports multiple ion species and the application of species-selective forces.
In this example we model the sympathetic cooling of \speciesNH{} ions by laser-cooled calcium ions \cite{Okada2010,Ostendorf2006}.

\begin{lstlisting}
sim = LAMMPSSimulation();
sim.GPUAccel = 1;
SetSimulationDomain(sim, 1e-3,1e-3,1e-3);

NH3 = AddAtomType(sim, 1, 17);
Ca40 = AddAtomType(sim, 1, 40);
createIonCloud(sim, 5e-4, NH3, 20);
createIonCloud(sim, 5e-4, Ca40, 50);

rf = 5.634e6;
sim.Add(linearPT(252.2, 5, 10e-3, 3.5e-3, 0.3, rf));

allBath = langevinBath(1e-3, 30e-6);
sim.Add(allBath);
sim.Add(evolve(100000));

sim.Add(dump('output.txt', {'id', 'x', 'y', 'z', timeAvg({'vx', 'vy', 'vz'}, 1/rf)}, 20));
sim.Add(evolve(60000));

sim.Remove(allBath);
sim.Add(laserCool(Ca40, [1e5 0 0]));
sim.Add(evolve(120000));

sim.Execute();
\end{lstlisting}

We start by defining the \code{sim} object and this time we set \code{GPUAccel} to enable GPU acceleration.
We create the two ion clouds, define the Paul trap parameters as before, and couple both species to a Langevin bath at \SI{1}{\milli\K}. 
The first \code{evolve} brings both species to thermal equilibrium with the bath.
At this time $t_\text{cool}$ we remove the bath with \code{sim.Remove}, add laser-cooling along the $\hat{x}$-axis to the \ca ions, and \code{evolve} the system again.

\begin{figure}[ht]
\centering
     \includegraphics[width=\columnwidth]{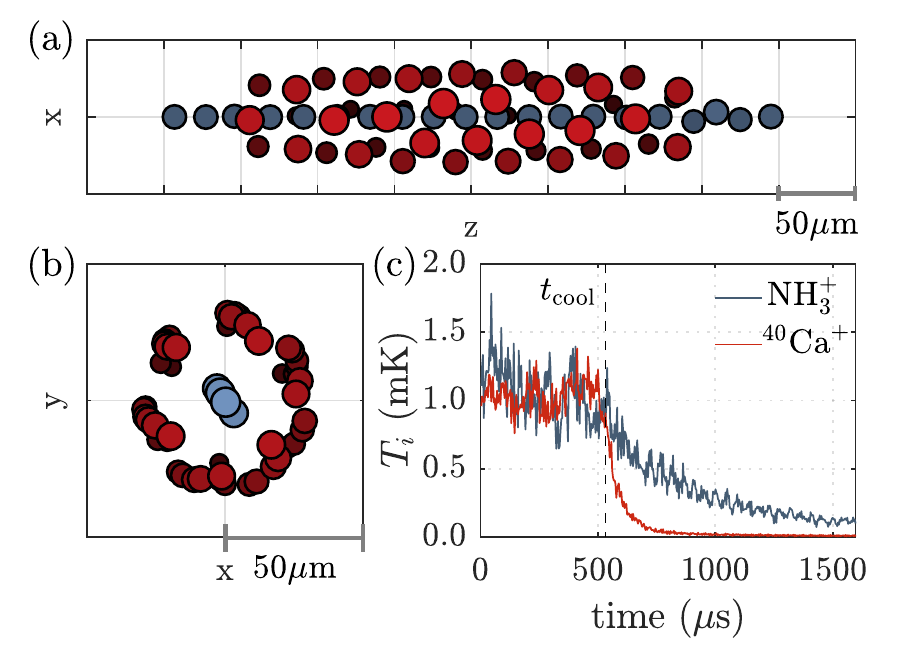}
\caption{A dual-species simulation of 20 \speciesNH{} ions and 50 \ca{} ions in a linear Paul trap, seen (a) from the side and (b) along the axis of the trap. The \speciesNH{} ions (blue) are confined tightly to the axis of the trap, while the lower charge-to-mass ratio \ca{} ions (red) form a sheath that surrounds them. c) At $t_\text{cool}$ (black, dashed line) a \SI{1}{\milli\K} Langevin bath is removed and laser cooling is applied to the $\hat{x}$-axis motion of the \ca{} ions. The temperature of the \ca{} ensemble is cooled quickly, while the interaction between the species causes sympathetic cooling of the \speciesNH{} ions at a slower rate.}
\label{fig:example2}
\end{figure}

\refform{fig:example2} shows the final equilibrium positions and the kinetic energy of both species.
The lighter \speciesNH{} ions are confined tightly in the centre of the trap while the heavier \ca{} ions form a sheath around them.
The energy of both species is damped over time; it takes longer for the \speciesNH{} ions to reach the equilibrium temperature since they are only indirectly cooled via interactions with the cloud of laser-cooled \ca{} ions.

\subsection{Charged rigid body}

This example simulates a group of ions which are bound together to create a rigid body. This body then interacts with a cloud of ions.

\begin{lstlisting}
sim = LAMMPSSimulation();
sim.SetSimulationDomain(1e-3,1e-3,1e-3);

mass = 40; charge = 1;
Ca = sim.AddAtomType(charge, mass);
createIonCloud(sim, 1e-4, Ca, 30);

rodz = (-2:0.5:2) * 5e-6;
rody = zeros(size(rodz));
rodx = zeros(size(rodz));
rodAtoms = placeAtoms(sim, Ca, rodx', rody', rodz');
rod = sim.Group(rodAtoms);
rod.Rigid = true;

RF = 3.85e6;
sim.Add(linearPT(500, 15, 2.75e-3, 3.5e-3, 0.3, RF));
sim.Add(langevinBath(0, 1e-5));

sim.Add(dump('pos.txt', {'id', 'x', 'y', 'z'}, 2));
secularVel = timeAvg({'vx', 'vy', 'vz'}, 1/RF)
sim.Add(dump('secV.txt', {'id', secularVel}));
sim.Add(evolve(5000));

sim.Execute();
\end{lstlisting}

\begin{figure}[ht]
\centering
     \includegraphics[width=\columnwidth]{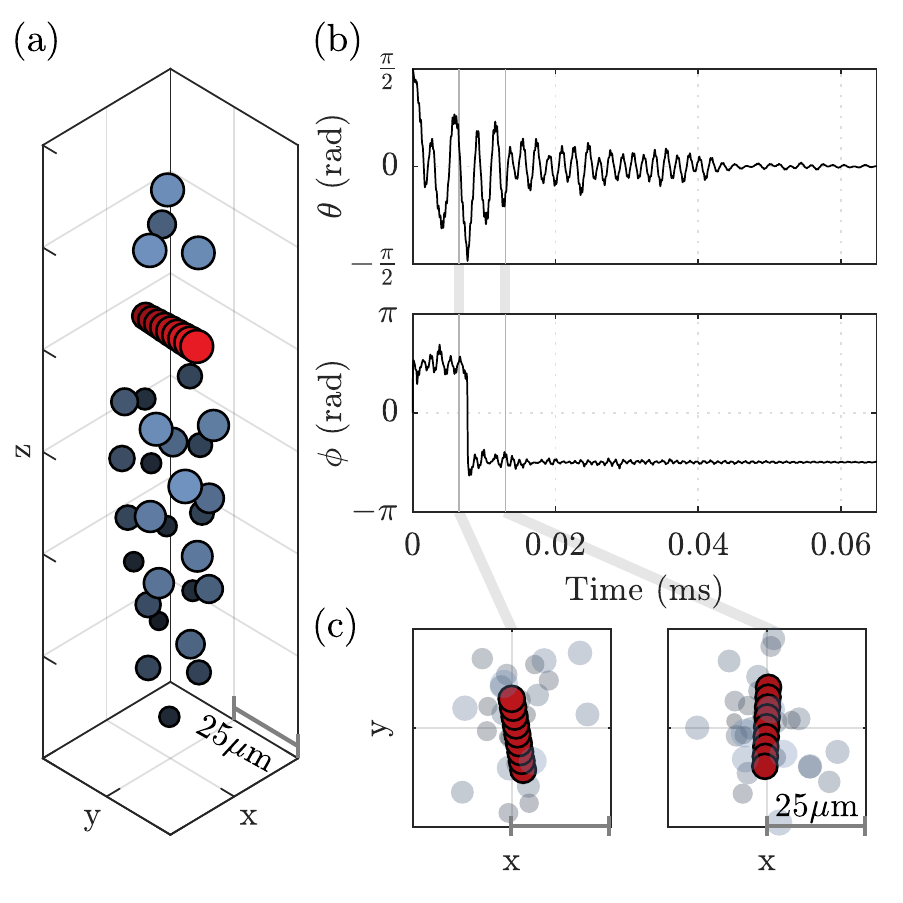}
\caption{
A cloud of \ca ions and a rigid rod, which is composed of 9 ions fixed together. After cooling via coupling to a Langevin bath, the ions and rod arrange into a Coulomb crystal.
(a) The final positions of ions in the cloud (blue) and the rod (red).
(b) The polar angles $\theta(t)$ and $\phi(t)$ describe the orientation of the rod. These quantities are plotted against time $t$, showing that the rod initially tumbles before aligning with respect to the trap and other ions and performing a small thermal motion.
(c) Two panels showing a top view of ion positions at early times in the simulation, before the Coulomb crystal has formed.
The plot shows the free ions as transparent so that the rod is clearly visible.
The specific times shown are illustrated by the vertical lines in panel b.
}
\label{fig:rigidbody}
\end{figure}

We start as before, configuring a simulation with two sets of calcium ions.
The first are placed randomly, while the second set of ions are arranged in a line and grouped together, forming a rod.
Setting the \code{Rigid} property of the group to true fixes the relative positions of ions.
The system is coupled to a zero temperature bath and forms a Coulomb crystal after time evolution.
\refform{fig:rigidbody} shows the results of the simulation.
Ions in the rod form a rigid body, which moves and rotates according to the forces applied to the constituent ions.

\subsection{Benchmarking}
\label{sec:GPU}

Our benchmark simulations are performed for different numbers of ions $N$ and using two standard workstations: a Dell Optiplex 9020 with an i7-4790 4-core CPU at \SI{3.6}{\giga\Hz} and 8 GB of RAM, and a Lenovo ThinkCentre M900 with an i7-6700 4-core CPU at \SI{3.4}{\giga\Hz} and 24 GB of RAM. In addition, the Lenovo is fitted with an nVidia Titan XP GPU, which has 12 GB of GDDR5 RAM and a processor frequency of \SI{1.5}{\giga\Hz}.

The computation time in the CPU-only simulations begins to scale as $N^2$ above $N\approx 30$ ions, where it becomes dominated by calculation of the long-ranged Coulomb repulsion.
The brute-force method used involves looping over interacting pairs, the total number of which is proportional to $N^2$.
For $N<30$ the step computation time is independent of ion number, indicating that an overhead in the \lammpsSoftware{} integration loop limits the speed of the calculation.

\begin{figure}[ht]
\centering
     \includegraphics[width=\columnwidth]{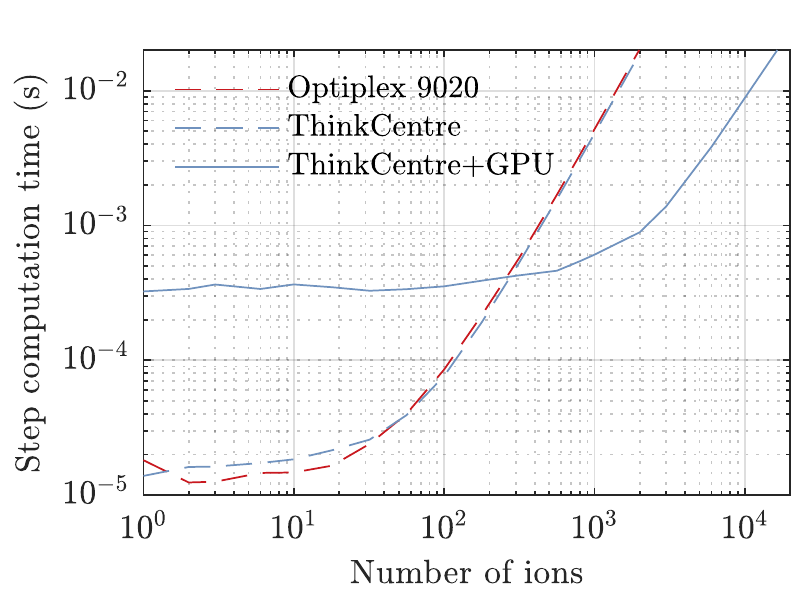}
\caption{Average computation time per timestep in the benchmark simulations, as a function of ion number. Dashed lines denote CPU-only simulations, solid lines show simulations that are accelerated by the GPU. See text for discussion and a description of the two workstations used.}
\label{fig:benchmarking}
\end{figure}

Setting \code{sim.GPUAccel} to true enables GPU acceleration.
This reduces the calculation time for pairwise interactions between large numbers of ions, and becomes advantageous when simulating more than a few hundred ions (see \refform{fig:benchmarking}).
GPU Acceleration corresponds to an order-of-magnitude improvement in time for $N\approx 1000$.
For $N<300$, GPU acceleration becomes inefficient because of the overhead required to configure tasks on the GPU.

\section{Verifications}
\label{sec:Validation}

\pl is distributed with a test suite which verifies functionality by comparing generated output to analytic results.
These validations provide a means to detect if modifications introduce errors, ensuring the integrity of \pl{} during further development.
This section describes each test simulation, the expected behaviour, and the functionality tested.

\subsection{Secular motion frequencies}

\begin{figure}[ht]
\centering
     \includegraphics[width=\columnwidth]{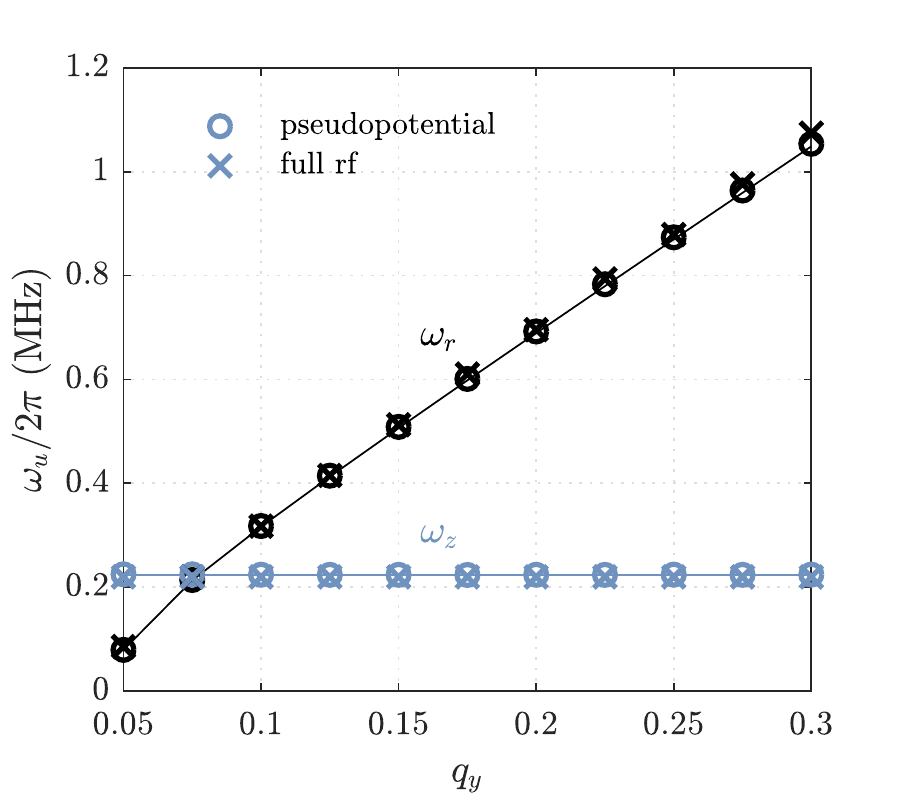}
\caption{
Secular frequencies $\omega_u / 2\pi$ measured from simulations of a single ion in a linear Paul trap.
Those in the radial direction (black) and along \coordAxis{z} (blue) are compared to theory (solid lines), for simulations using either the pseudopotential (circles) or full electric fields (crosses).
}
\label{fig:ver1}
\end{figure}

\code{SecularFrequencies\_pseudoPot.m} and \code{\_fullRF.m} test the \pl implementation of the linear Paul trap, using either the pseudopotential approximation or the full electric fields respectively.
These verifications configure a linear Paul trap, simulate the motion of a single ion and compare the single ion oscillation frequencies to those predicted by eq.~\eqref{eq:pseudopot} (see \refform{fig:ver1}).
We take the measured oscillation frequency along each axis to be the Fourier component with the largest amplitude.

\subsection{Equilibrium separation}

The lowest energy configuration when $\omega_{x,y} \gg \omega_z$ is a linear Coulomb crystal of $N$ ions aligned along the axis $\hat{e}_z$.
The position of the $i$th ion, $z_i$, depends on the interplay between the confinement along the $\hat{e}_z$-axis, which compresses the chain, and the Coulomb repulsion from the other ions.
These positions may be found by minimising the potential energy of \refform{eq:fullPotential}.
The separation between neighbouring ions increases away from the centre of the chain, where the minimum separation is approximately~\cite{James1998}
\begin{equation}
    \frac{z_{\mathrm{min}}}{l} = \frac{2.018}{N^{0.559}}, \quad \text{with}\  l^3 = \frac{Q^2}{4\pi\epsilon_0M\omega_z^2}.
\end{equation}

\begin{figure}[ht]
\centering
     \includegraphics[width=\columnwidth]{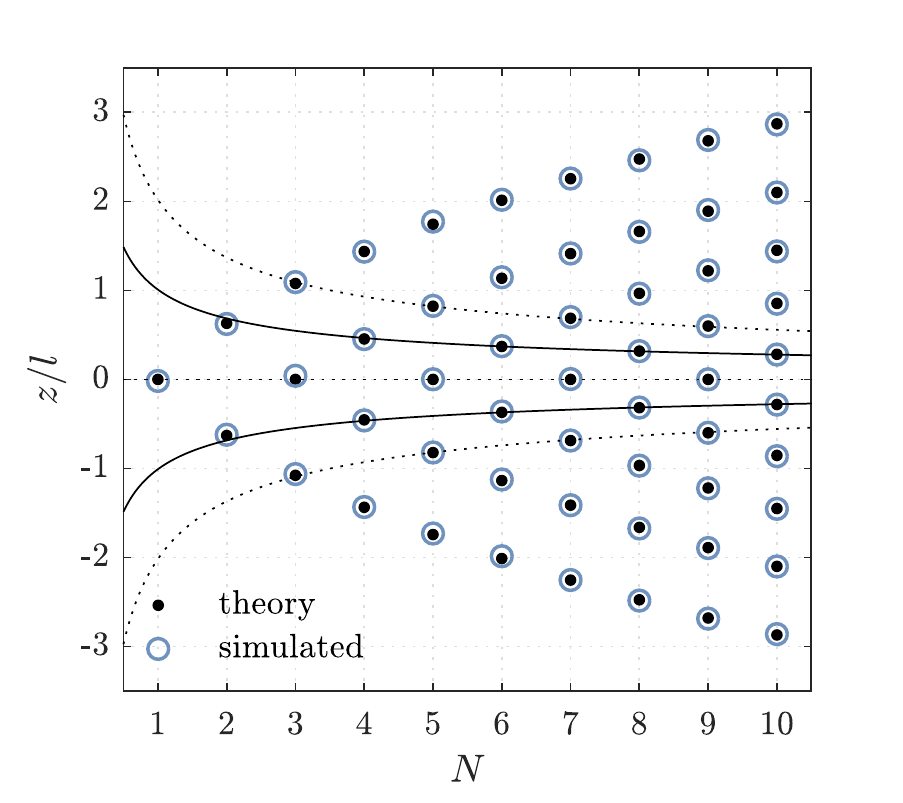}
\caption{
Normalised equilibrium positions $z/l$ of $N$ ions in a linear chain.
There is good agreement between theoretical predictions (black dots) and simulation results (blue circles) performed using the pseudopotential approximation.
For even $N>1$, the two neighbouring ions that are closest are those at $z=\pm z_\mathrm{min}/2$ (solid black lines).
For odd $N>1$, they are the pairs positioned at $z=0$ and $z=\pm z_\mathrm{min}$ (dotted black lines).
}
\label{fig:ver_equilPos}
\end{figure}

\code{AxialSeparation.m} minimises the energy of ions arranged in a linear Coulomb crystal, and compares the simulated positions $z_i$ to those predicted by theory~\cite{James1998} (see \refform{fig:ver_equilPos}).
Simulations of these equilibrium positions test the implementations of both the Coulomb interaction and the axial confinement strength of the Paul trap.

\subsection{Normal modes of motion}

\code{NormalModes\_Linear.m} simulates the normal mode spectrum for a fixed number of ions in a Coulomb crystal.
First it drives the system to its lowest energy configuration, then allows it to evolve without any damping using the pseudopotential approximation.
The normal modes are excited by delivering a periodic kick to the ensemble, which is implemented by randomising the velocities of the ions.
The kicks are broadband enough to excite all the normal modes in the Coulomb crystal, and are sufficiently small in amplitude that the crystal does not `melt'.
The positions of the ions are written to a file twice for every \rf{} cycle, so that all the modes are visible in the final spectrum, and the amplitudes of these modes are calculated by Fourier transforming the trajectories.

\begin{figure}[ht]
\centering
     \includegraphics[width=\columnwidth]{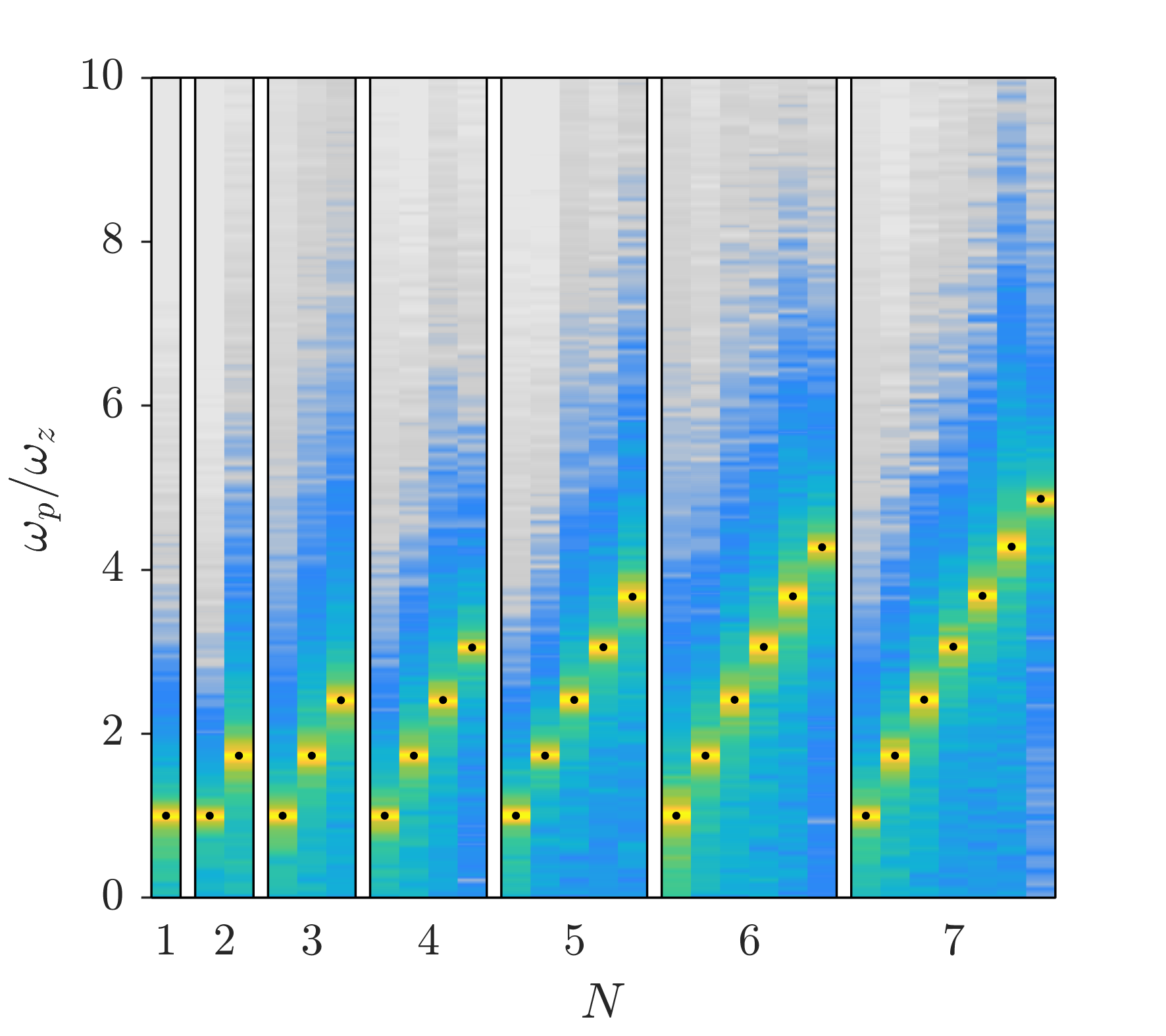}
\caption{
The spectral decomposition of each axial normal mode is shown for simulations of $N$-ion linear Coulomb crystals, where $N \in \{1,7\}$.
The vertical axis denotes frequency, normalised with respect to $\omega_z$.
The frequency spectrum for each normal mode is shown as a vertical colored stripe; the logarithm of intensities are plotted using a colour map where yellow depicts a maximum and grey depicts vanishing amplitude.
The maximum amplitude of each mode is in agreement with the predicted frequency of that normal mode (black dot).
}
\label{fig:ver_normalModes_linear}
\end{figure}

The frequencies of the normal modes for a chain of $N$ ions are the eigenvalues of the Hessian matrix~\cite{kielpinskiSympatheticCoolingTrapped2000} constructed from the spatial derivatives of \refform{eq:fullPotential}.
For small numbers of ions ($N<7$) the theoretical values are tabulated in Ref.~\cite{James1998}. 
\refform{fig:ver_normalModes_linear} shows that the theoretical normal modes are in agreement with those calculated by \pl.

\subsection{Cooling mechanisms}

\code{LaserCooling.m} simulates the trajectories of free, non-interacting ions that are strongly laser-cooled along a single randomly-chosen direction, with a characteristic damping time $\tau=\SI{1}{\micro s}$.
It then projects the velocity of each ion onto the laser-cooling axis, fits them with an exponential function and compares the result of the fit to $\tau$.
The perpendicular velocity components are unaffected by the laser-cooling force.

\begin{figure}[ht]
\centering
     \includegraphics[width=\columnwidth]{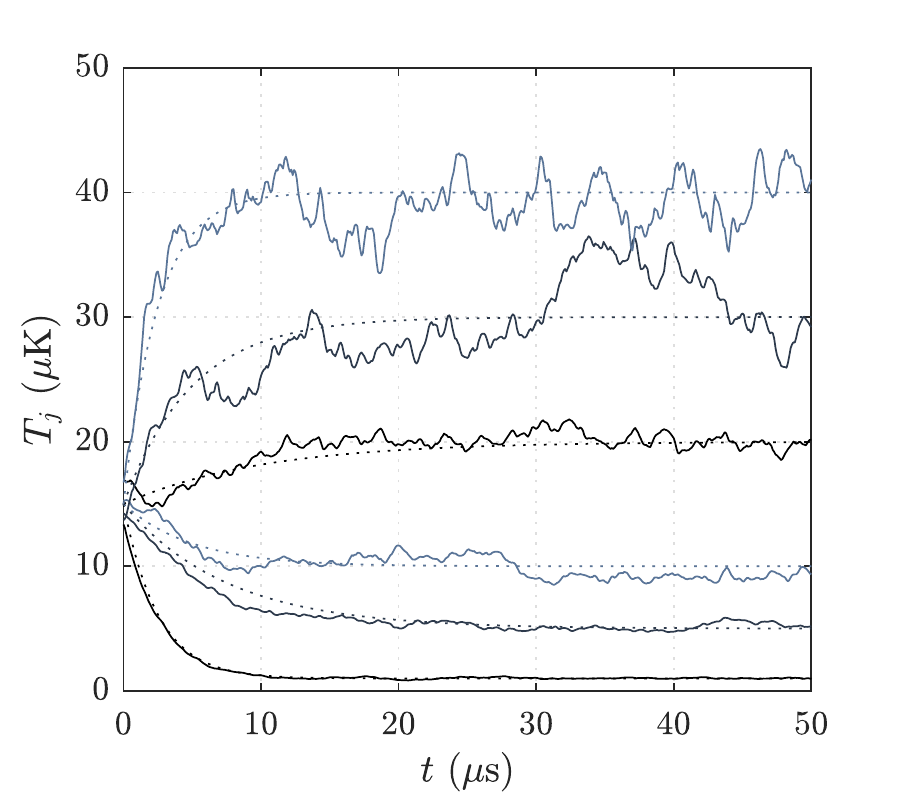}
\caption{
Groups of free, non-interacting atoms are each coupled to a different Langevin bath and the dynamics simulated.
Measured temperatures $T_j$ of each group are shown as solid lines, while dotted lines show the predicted temperature of that group, given the known coupling strength and temperature $\Phi_j$ of the associated bath.
}
\label{fig:ver_langevin}
\end{figure}

\code{LangevinBath.m} simulates groups of free, non-interacting atoms. 
Each group $j$ is coupled to a distinct Langevin bath with time constant $c_j$ and temperature $\Phi_j$. 
The equipartition theorem relates the temperature $T_j$ of each group to its kinetic energy, $3 k_B T_j = M \langle v_x^2 + v_y^2 + v_z^2 \rangle$, where $k_B$ is the Boltzmann constant. 
\refform{fig:ver_langevin} shows $T_j$ for each group as a function of time, along with theoretical curves (no fit) showing the predicted temperature as each ensemble thermalizes with the associated bath in accordance with Newton's law of cooling.

\section{Implementation Details}
\label{sec:Implementation}

Having described the usage and validity of our package, we now address its implementation.
Although an understanding of LAMMPS is not required to use \pl, in this section we provide pertinent details of the \lammpsSoftware{} configuration and features used, which are useful for implementing new features.
For clarity, the commands specific to LAMMPS are indicated \lammps{like so}.

\subsection{Input file preamble}

The input file starts with a declaration of the static properties of the simulation. 
All quantities are defined in terms of SI units using \lammps{units si}.
The \lammps{package gpu} and \lammps{suffix gpu} commands toggle GPU acceleration on/off by overriding fix\footnote{a \emph{fix} in \lammpsSoftware{} refers to an operation applied during integration.} styles in \lammpsSoftware{} with equivalent GPU implementations~\cite{Brown2011,Brown2012,Brown2013}. 
The \lammps{boundary}, \lammps{region} and \lammps{create\_box} commands define a non-periodic simulation region. 

\subsection{Ions in LAMMPS}

The LAMMPS \lammps{atom\_style charge} command configures the simulation to represent ions as atoms with a velocity, position, id and atom type.
The \lammps{create\_atoms} command adds atoms one-by-one to the input file, placing a single atom of specified type and position each time.
The \lammps{mass} and \lammps{set type} commands subsequently configure the different atom types.

\subsection{Interactions}
\label{sec:pairwise}

To determine which atoms should interact through pairwise potentials, LAMMPS creates neighbour lists of atoms within a threshold distance of each other. 
The threshold is equal to the force cut-off distance plus a \textit{skin} distance which for greater values reduces the frequency at which the list becomes invalid and must be rebuilt. 
This approach is computationally efficient for systems of short-ranged forces in which atoms frequently aquire new neighbours, but is redundant for systems of ions because each ion interacts with all other ions for the duration of the simulation. 
As such, \pl configures LAMMPS to build the neighbour list once at the start of the simulation using the \lammps{nsq} style, and with all ions listed as neighbours.

The \lammps{pair\_style} command specifies the interaction model used between neighbouring atoms. 
\pl disables short-ranged interactions such as Van der Waals forces because the spatial separation between ions is typically orders of magnitude larger than the length scales of these forces. 
The pair style \lammps{coul/cut} defines a truncated Coulomb interaction. 
The default cut-off distance is chosen to be \SI{10}{m}, which makes it effectively infinite for typical-size systems.

The pairwise interaction represents the most computationally intense part of the integration. 
A brute force method that directly sums these forces scales as $\mathcal O(N^2)$, which becomes unfavourable for large particle numbers, but is well-suited to parallel computation.
Other methods for calculating long-ranged forces are implemented in \lammpsSoftware{}, such as multi-level summation~\cite{Hardy2009}. Ewald summation and particle-particle/particle-mesh methods are also available for systems with periodic boundary conditions~\cite{Brown2012}, although these are not suited to ion traps.
We have found that \pl{} performs well using the brute-force summation method for our typical simulation sizes of 100 atoms (see \refform{sec:GPU}), and so have not exposed these other methods to \pl{}.

\subsection{Simulation elements}

The remainder of the input file describes successive definitions of forces, variables, output, and evolution.
These are each contained in the ordered array \code{sim.Elements} as \code{InputFileElement}s.
\pl enumerates the array, generating the required input file content by invoking each element's \code{createInputFileText} function.
This is also the mechanism by which \pl can be extended; the extension only needs to implement the \code{createInputFileText} method to return the appropriate input file text required to configure the targeted LAMMPS feature.
For instance, a \pl{} \code{efield} element inserts the text \lammps{fix <fixID> all efield <Ex> <Ey> <Ez>} to the input file, using a unique identifier \code{<fixID>} for the force and specifying the components of the electric field.

\subsection{Integration}

The \lammps{nve} integrator updates the positions of single atoms at each integration step according to the Newtonian equations of motion.
We work in the NVE ensemble in which the system is isolated and the total number of ions is conserved.
The \lammps{rigid} integrator calculates the motion of rigid bodies, maintaining the relative position of their constituent atoms.
\lammps{timestep} defines the fixed step duration used for time integration.
\pl configures LAMMPS to print diagnostic information about compute resource usage every 10000 steps using the \lammps{thermo} command.

To minimise the energy of an ensemble the \lammps{minimize} and \lammps{min\_style quickmin} commands offer alternative integration schemes that converge in a smaller number of steps (see \refform{sec:Minimisation}).

\section{Conclusion}
\label{sec:Conclusion}

\pl allows full configuration and execution of ion trap simulations from within a Matlab or Python environment. 
It offloads the computationally intensive part to \lammpsSoftware{}, while the environment of choice is used for configuration and analysis. 
No knowledge of \lammpsSoftware{}' specialist command language is required. 

Molecular dynamics codes such as \lammpsSoftware{} are typically associated with calculation of short-range inter-atomic forces that drive mesoscopic material and biophysical processes. 
We have shown that \lammpsSoftware{} can also efficiently model systems where long range interactions are dominant, such as electrodynamically confined clouds of ions.
Wrapping \lammpsSoftware{} brings a number of features to \pl, \eg support for rigidly bound clusters of ions, and GPU acceleration. 
\pl complements \lammpsSoftware{} by providing high-level specialised functions that simplify the description of the ion trap system.

All functionality of \pl has been verified by a thorough comparison to theory.
We have used \pl to simulate a range of single/multiple species and multi-frequency systems~\cite{footTwofrequencyOperationPaul2018,trypogeorgosCotrappingDifferentSpecies2016}.
\lammpsSoftware{} is a mature code with many other features that can be implemented in \pl.
These include electric fields defined by finite element methods (using the \lammps{atc} package), and ion-neutral interactions using direct-simulation Monte Carlo methods.
These future extensions will allow \pl to deal with even more complex scenarios, such as: the cooling of large, levitated objects via interaction with a buffer gas or feedback~\cite{nagornykhParametricStabilizationCooling2014}; rarefied gas dynamics to simulate ion collisions with a buffer gas~\cite{birdMolecularGasDynamics1994}; coarse-grained molecular dynamics simulations of protein structures in aqueous Paul traps~\cite{guanPaulTrappingCharged2011,josephLongDNASegment2010}; studies of cold molecule production/reactions~\cite{bellUltracoldMoleculesUltracold2009}; and interactions of ions with polar molecules~\cite{idziaszekIonassistedGroundstateCooling2011}.

\section{Getting Started}
\label{sec:gettingStarted}

Up-to-date stable releases of LAMMPS are available at \texttt{http://lammps.sandia.gov/}.
Multiple build options for LAMMPS exist. 
At least the \lammps{misc} package must be included, and optionally the \lammps{rigid} and \lammps{gpu}\footnote{The \lammps{gpu} package is included by default in all windows binaries.} packages for rigid-body and GPU support respectively.
The most recent version of \pl{}'s MATLAB distribution is available at \texttt{bitbucket.org/footgroup/lion.git}, and further documentation and examples can be found in the source.
For instructions to install and configure \pl, please see the readme file. 
The Python version of \pl can be downloaded from \texttt{bitbucket.org/dtrypogeorgos/pylion}, along with documentation explaining how to get started.

\section*{Acknowledgements}

EB acknowledges support from a Doctoral Training Studentship funded by the EPSRC.
The authors thank Tiffany Harte and Michal Hejduk for comments on the manuscript.

\bibliographystyle{elsarticle-num}
\bibliography{refs}

\begin{thebibliography}{10}
\expandafter\ifx\csname url\endcsname\relax
  \def\url#1{\texttt{#1}}\fi
\expandafter\ifx\csname urlprefix\endcsname\relax\def\urlprefix{URL }\fi
\expandafter\ifx\csname href\endcsname\relax
  \def\href#1#2{#2} \def\path#1{#1}\fi

\bibitem{Paul1990}
W.~Paul, Electromagnetic traps for charged and neutral particles, Rev. Mod.
  Phys. 62 (1990) 531--540.
\newblock \href {https://doi.org/10.1103/RevModPhys.62.531}
  {\path{doi:10.1103/RevModPhys.62.531}}.

\bibitem{marchQuadrupoleIonTrap2005}
Quadrupole {{Ion Trap Mass Spectrometry}}, {{Volume}} 165, {{Second Edition}},
  {Wiley-VCH}.

\bibitem{staffordRecentImprovementsAnalytical1984}
G.~C. Stafford, P.~E. Kelley, J.~E.~P. Syka, W.~E. Reynolds, J.~F.~J. Todd,
  Recent improvements in and analytical applications of advanced ion trap
  technology 60~(1)  85--98.
\newblock \href {https://doi.org/10.1016/0168-1176(84)80077-4}
  {\path{doi:10.1016/0168-1176(84)80077-4}}.

\bibitem{IonTraps1996}
Ion {{Traps}}, International {{Series}} of {{Monographs}} on {{Physics}},
  {Oxford University Press}.

\bibitem{benhelmFaulttolerantQuantumComputing2008}
J.~Benhelm, G.~Kirchmair, C.~F. Roos, R.~Blatt, Towards fault-tolerant quantum
  computing with trapped ions 4~(6)  463--466.
\newblock \href {https://doi.org/10.1038/nphys961}
  {\path{doi:10.1038/nphys961}}.

\bibitem{vangorpSimbucaUsingGraphics2011}
S.~Van~Gorp, M.~Beck, M.~Breitenfeldt, V.~De~Leebeeck, P.~Friedag, A.~Herlert,
  T.~Iitaka, J.~Mader, V.~Kozlov, S.~Roccia, G.~Soti, M.~Tandecki, E.~Traykov,
  F.~Wauters, C.~Weinheimer, D.~Zákoucký, N.~Severijns, Simbuca, using a
  graphics card to simulate {{Coulomb}} interactions in a penning trap 638~(1)
  192--200.
\newblock \href {https://doi.org/10.1016/j.nima.2010.11.032}
  {\path{doi:10.1016/j.nima.2010.11.032}}.

\bibitem{dahlSIMIONPCPS21990}
D.~A. Dahl, J.~E. Delmore, A.~D. Appelhans, {{SIMION PC}}/{{PS2}} electrostatic
  lens design program 61~(1)  607--609.
\newblock \href {https://doi.org/10.1063/1.1141932}
  {\path{doi:10.1063/1.1141932}}.

\bibitem{schillerMolecularDynamicsSimulation2003}
S.~Schiller, C.~Lammerzahl, Molecular dynamics simulation of sympathetic
  crystallization of molecular ions 68~(5)  053406.
\newblock \href {https://doi.org/10.1103/PhysRevA.68.053406}
  {\path{doi:10.1103/PhysRevA.68.053406}}.

\bibitem{trottGeneralPurposeMolecular2010}
C.~Trott, L.~Winterfeld, General purpose {{Molecular Dynamics Simulations}} on
  {{GPUs}}: {{Issues}} of {{Pair Forces}} and {{Scaling}} to large
  {{Clusters}}\href {http://arxiv.org/abs/1009.4330} {\path{arXiv:1009.4330}}.

\bibitem{mattheyPROTOMOLObjectOrientedFramework2002}
T.~Matthey, T.~Cickovski, S.~Hampton, A.~Ko, Q.~Ma, {{PROTOMOL}}, an
  {{Object}}-{{Oriented Framework}} for {{Prototyping Novel Algorithms}} for
  {{Molecular Dynamics}}, in: In {{Computational Science}}—{{ICCS}} 2003,
  {{International Conference}}, {and St}.

\bibitem{NAMD}
J.~C. Phillips, R.~Braun, W.~Wang, J.~Gumbart, E.~Tajkhorshid, E.~Villa,
  C.~Chipot, R.~D. Skeel, L.~Kalé, K.~Schulten, Scalable molecular dynamics
  with namd, Journal of Computational Chemistry 26~(16)  1781--1802.
\newblock \href
  {http://arxiv.org/abs/https://onlinelibrary.wiley.com/doi/pdf/10.1002/jcc.20289}
  {\path{arXiv:https://onlinelibrary.wiley.com/doi/pdf/10.1002/jcc.20289}},
  \href {https://doi.org/10.1002/jcc.20289} {\path{doi:10.1002/jcc.20289}}.

\bibitem{Gromacs}
H.~Berendsen, D.~van~der Spoel, R.~van Drunen, {GROMACS}: A message-passing
  parallel molecular dynamics implementation, Computer Physics Communications
  91~(1) (1995) 43 -- 56.
\newblock \href {https://doi.org/https://doi.org/10.1016/0010-4655(95)00042-E}
  {\path{doi:https://doi.org/10.1016/0010-4655(95)00042-E}}.

\bibitem{Hesse2002}
E.~Hesse, Z.~Ulanowski, P.~Kaye, Stability characteristics of cylindrical
  fibres in an electrodynamic balance designed for single particle
  investigation, Journal of Aerosol Science 33.
\newblock \href {https://doi.org/10.1016/S0021-8502(01)00153-7}
  {\path{doi:10.1016/S0021-8502(01)00153-7}}.

\bibitem{kaneLevitatedSpinningGraphene2010a}
B.~E. Kane, Levitated spinning graphene flakes in an electric quadrupole ion
  trap 82~(11)  115441.
\newblock \href {https://doi.org/10.1103/PhysRevB.82.115441}
  {\path{doi:10.1103/PhysRevB.82.115441}}.

\bibitem{Delord2017}
T.~{Delord}, L.~{Nicolas}, Y.~{Chassagneux}, G.~{H{\'e}tet}, Strong coupling
  between a single nitrogen-vacancy spin and the rotational mode of diamonds
  levitating in an ion trap, Phys.~Rev.~A 96~(6) (2017) 063810.
\newblock \href {http://arxiv.org/abs/1702.00774} {\path{arXiv:1702.00774}},
  \href {https://doi.org/10.1103/PhysRevA.96.063810}
  {\path{doi:10.1103/PhysRevA.96.063810}}.

\bibitem{footAtomicPhysics2004}
C.~J. Foot, Atomic {{Physics}}, {OUP Oxford}.

\bibitem{berkelandMinimizationIonMicromotion1998}
D.~J. Berkeland, J.~D. Miller, J.~C. Bergquist, W.~M. Itano, D.~J. Wineland,
  Minimization of ion micromotion in a {{Paul}} trap 83~(10)  5025--5033.
\newblock \href {https://doi.org/10.1063/1.367318}
  {\path{doi:10.1063/1.367318}}.

\bibitem{willitschChemicalApplicationsLaser2008}
S.~Willitsch, M.~T. Bell, A.~D. Gingell, T.~P. Softley, Chemical applications
  of laser- and sympathetically-cooled ions in ion traps 10~(48)  7200--7210.
\newblock \href {https://doi.org/10.1039/B813408C}
  {\path{doi:10.1039/B813408C}}.

\bibitem{zhangMoleculardynamicsSimulationsCold2007a}
C.~B. Zhang, D.~Offenberg, B.~Roth, M.~A. Wilson, S.~Schiller,
  Molecular-dynamics simulations of cold single-species and multispecies ion
  ensembles in a linear {{Paul}} trap 76~(1)  012719.
\newblock \href {https://doi.org/10.1103/PhysRevA.76.012719}
  {\path{doi:10.1103/PhysRevA.76.012719}}.

\bibitem{nasseInfluenceBackgroundPressure2001}
M.~Nasse, C.~Foot, Influence of background pressure on the stability region of
  a {{Paul}} trap 22  563--573.
\newblock \href {https://doi.org/10.1088/0143-0807/22/6/301}
  {\path{doi:10.1088/0143-0807/22/6/301}}.

\bibitem{hasegawaDynamicsSingleParticle1995}
T.~Hasegawa, K.~Uehara, Dynamics of a single particle in a {{Paul}} trap in the
  presence of the damping force 61  159--163.
\newblock \href {https://doi.org/10.1007/BF01090937}
  {\path{doi:10.1007/BF01090937}}.

\bibitem{LammpsManual}
{LAMMPS} user manual, \url{https://lammps.sandia.gov/doc/Manual.html},
  accessed: 2019-05-29.

\bibitem{Okada2010}
K.~Okada, M.~Wada, T.~Takayanagi, S.~Ohtani, H.~A. Schuessler, Characterization
  of ion coulomb crystals in a linear paul trap, Phys. Rev. A 81 (2010) 013420.
\newblock \href {https://doi.org/10.1103/PhysRevA.81.013420}
  {\path{doi:10.1103/PhysRevA.81.013420}}.

\bibitem{Ostendorf2006}
A.~Ostendorf, C.~B. Zhang, M.~A. Wilson, D.~Offenberg, B.~Roth, S.~Schiller,
  Sympathetic cooling of complex molecular ions to millikelvin temperatures,
  Phys. Rev. Lett. 97 (2006) 243005.
\newblock \href {https://doi.org/10.1103/PhysRevLett.97.243005}
  {\path{doi:10.1103/PhysRevLett.97.243005}}.

\bibitem{James1998}
D.~James, Quantum dynamics of cold trapped ions with application to quantum
  computation, Applied Physics B 66~(2) (1998) 181--190.
\newblock \href {https://doi.org/10.1007/s003400050373}
  {\path{doi:10.1007/s003400050373}}.

\bibitem{kielpinskiSympatheticCoolingTrapped2000}
D.~Kielpinski, B.~E. King, C.~J. Myatt, C.~A. Sackett, Q.~A. Turchette, W.~M.
  Itano, C.~Monroe, D.~J. Wineland, W.~H. Zurek, Sympathetic cooling of trapped
  ions for quantum logic 61~(3)  032310.
\newblock \href {https://doi.org/10.1103/PhysRevA.61.032310}
  {\path{doi:10.1103/PhysRevA.61.032310}}.

\bibitem{Brown2011}
S.~J. P. A. N.~T. W.~M.~Brown, P.~Wang, Implementing molecular dynamics on
  hybrid high performance computers - short range forces, Comp.~Phys.~Comm. 182
  (2011) 898--911.

\bibitem{Brown2012}
W.~M. Brown, A.~Kohlmeyer, S.~J. Plimpton, A.~N. Tharrington, Implementing
  molecular dynamics on hybrid high performance computers –
  particle–particle particle-mesh, Computer Physics Communications 183~(3)
  (2012) 449 -- 459.
\newblock \href {https://doi.org/https://doi.org/10.1016/j.cpc.2011.10.012}
  {\path{doi:https://doi.org/10.1016/j.cpc.2011.10.012}}.

\bibitem{Brown2013}
Y.~M. W.~M.~Brown, Implementing molecular dynamics on hybrid high performance
  computers – three-body potentials, Comp.~Phys.~Comm. 184 (2013) 2785--2793.

\bibitem{Hardy2009}
D.~J. Hardy, J.~E. Stone, K.~Schulten, Multilevel summation of electrostatic
  potentials using graphics processing units, Parallel Computing 35~(3) (2009)
  164 -- 177, revolutionary Technologies for Acceleration of Emerging Petascale
  Applications.
\newblock \href {https://doi.org/https://doi.org/10.1016/j.parco.2008.12.005}
  {\path{doi:https://doi.org/10.1016/j.parco.2008.12.005}}.

\bibitem{footTwofrequencyOperationPaul2018}
C.~J. Foot, D.~Trypogeorgos, E.~Bentine, A.~Gardner, M.~Keller, Two-frequency
  operation of a {{Paul}} trap to optimise confinement of two species of ions
  430  117--125.
\newblock \href {https://doi.org/10.1016/j.ijms.2018.05.007}
  {\path{doi:10.1016/j.ijms.2018.05.007}}.

\bibitem{trypogeorgosCotrappingDifferentSpecies2016}
D.~Trypogeorgos, C.~J. Foot, Cotrapping different species in ion traps using
  multiple radio frequencies 94~(2)  023609.
\newblock \href {https://doi.org/10.1103/PhysRevA.94.023609}
  {\path{doi:10.1103/PhysRevA.94.023609}}.

\bibitem{nagornykhParametricStabilizationCooling2014}
P.~Nagornykh, B.~E. Kane, Parametric stabilization and cooling of
  microparticles in a quadrupole ion trap, in: Optical {{Trapping}} and
  {{Optical Micromanipulation XI}}, Vol. 9164, {International Society for
  Optics and Photonics}, p. 916405.
\newblock \href {https://doi.org/10.1117/12.2064969}
  {\path{doi:10.1117/12.2064969}}.

\bibitem{birdMolecularGasDynamics1994}
G.~A. Bird, Molecular {{Gas Dynamics}} and the {{Direct Simulation}} of {{Gas
  Flows}}, 2nd Edition, {Oxford University Press, USA}.

\bibitem{guanPaulTrappingCharged2011}
W.~Guan, S.~Joseph, J.~H. Park, P.~S. Krstic, M.~A. Reed, Paul trapping of
  charged particles in aqueous solution 108~(23)  9326--9330.
\newblock \href {http://arxiv.org/abs/21606331} {\path{arXiv:21606331}}, \href
  {https://doi.org/10.1073/pnas.1100977108}
  {\path{doi:10.1073/pnas.1100977108}}.

\bibitem{josephLongDNASegment2010}
S.~Joseph, W.~Guan, M.~A. Reed, P.~S. Krstic, Long {{DNA}} segment in a linear
  nanoscale {{Paul}} trap 21~(1)  015103.
\newblock \href {http://arxiv.org/abs/19946172} {\path{arXiv:19946172}}, \href
  {https://doi.org/10.1088/0957-4484/21/1/015103}
  {\path{doi:10.1088/0957-4484/21/1/015103}}.

\bibitem{bellUltracoldMoleculesUltracold2009}
M.~T. Bell, T.~P. Softley, Ultracold molecules and ultracold chemistry 107~(2)
  99--132.
\newblock \href {https://doi.org/10.1080/00268970902724955}
  {\path{doi:10.1080/00268970902724955}}.

\bibitem{idziaszekIonassistedGroundstateCooling2011}
Z.~Idziaszek, T.~Calarco, P.~Zoller, Ion-assisted ground-state cooling of a
  trapped polar molecule 83~(5)  053413.
\newblock \href {https://doi.org/10.1103/PhysRevA.83.053413}
  {\path{doi:10.1103/PhysRevA.83.053413}}.

\end{thebibliography}

\end{document}